\documentclass[12pt]{article}

\usepackage{graphicx}	
\usepackage{amsbsy}		
\usepackage{amsfonts}		
\usepackage{amsmath}		
\usepackage{amssymb}		
\usepackage{wasysym}
\usepackage{multirow}
\usepackage{psfrag}
\usepackage{url}
\usepackage{cite}

\usepackage{epsfig,latexsym,cancel}

\newcommand{\beq}{\begin{eqnarray}}
\newcommand{\eeq}{\end{eqnarray}}

\newcommand{\ba}{\begin{array}}
\newcommand{\ea}{\end{array}}

\long\def\symbolfootnote[#1]#2{\begingroup%
\def\thefootnote{\fnsymbol{footnote}}\footnote[#1]{#2}\endgroup}

\newcommand{\no}{\nonumber}
\def\lsim{\mathrel{\rlap{\lower4pt\hbox{\hskip1pt$\sim$}}
    \raise1pt\hbox{$<$}}}         
\def\gsim{\mathrel{\rlap{\lower4pt\hbox{\hskip1pt$\sim$}}
    \raise1pt\hbox{$>$}}}         

\begin{document}

\hfill {\tt Bonn-Th-2011-10}

\hfill {\tt CERN-PH-TH/2011-093}

\hfill {\tt PI/UAN-2011-483FT}

\def\thefootnote{\fnsymbol{footnote}}

\begin{center}
\Large\bf\boldmath
\vspace*{0.8cm} Flavour physics constraints in the BMSSM
\unboldmath
\end{center}
\vspace{0.6cm}
\begin{center}
N.~Bernal$^{1,}$\footnote{Electronic address: nicolas@th.physik.uni-bonn.de},
M.~Losada$^{2,}$\footnote{Electronic address: malosada@uan.edu.co}, 
F.~Mahmoudi$^{3,4,}$\footnote{Electronic address: mahmoudi@in2p3.fr}\\[0.4cm] 
\vspace{0.6cm}
{\sl $^1$ Physikalisches Institut and Bethe Center for Theoretical Physics,\\ Universit\"at Bonn,
Nu\ss allee 12, D53115 Bonn, Germany}\\[0.4cm]
{\sl $^2$ Centro de Investigaciones, Cra 3 Este No 47A-15,\\ Universidad Antonio Nari\~{n}o, Bogot\'a, Colombia}\\[0.4cm]
{\sl $^3$ CERN Theory Division, Physics Department\\ CH-1211 Geneva 23, Switzerland}\\[0.4cm]
{\sl $^4$ Clermont Universit{\'e}, Universit\'e Blaise Pascal, CNRS/IN2P3,\\
LPC, BP 10448, 63000 Clermont-Ferrand, France}
\end{center}

\renewcommand{\thefootnote}{\arabic{footnote}}
\setcounter{footnote}{0}

\vspace{0.6cm}
\begin{abstract}
We study the implications of the presence of the two leading-order, non-renormalizable operators in the Higgs sector of the MSSM to flavour physics observables. We identify the constraints of flavour physics on the parameters of the BMSSM when we: a) focus on a region of parameters for which electroweak baryogenesis is feasible, b) use a CMSSM-like parametrization, and c) consider the case of a generic NUHM-type model. We find significant differences as compared to the standard MSSM case.
\end{abstract}

\newpage

\section{Introduction}
\label{sec:intro}

The Minimal Supersymmetric Standard Model (MSSM) is one of the most studied
models that probes effects from beyond the Standard Model (SM) physics.
Although the MSSM can provide many interesting features to
physics beyond the SM, there are severe constraints on the parameters when 
trying to provide a feasible scenario for
a Higgs boson mass consistent with experimental constraints,
electroweak baryogenesis (EWBG),
dark matter,
and flavour physics observables.

Within the MSSM, the tree level bound on the lightest Higgs mass
is violated, and hence significant corrections
arising from loops of top quarks and squarks are necessary.
But in order for these effects account for the Higgs mass, the top squarks must be quite massive
or the top squarks must be highly mixed.
This suggests that {\it if} low energy supersymmetry is important to the solution of the hierarchy
problem, there are likely to be additional degrees of freedom in the theory beyond
those of the MSSM.
Recently there has been a great interest in extensions of the MSSM by higher-dimension
operators~\cite{Strumia:1999jm,Casas:2003jx,Brignole:2003cm,PospelovRitz,Dine:2007xi,Antoniadis,Batra:2008rc}.
These may have an important impact on the Higgs sector, alleviating in particular the tension
in the MSSM that results from the LEP Higgs bounds, {\it i.e.} the so-called little hierarchy problem.
In particular, an attractive extension of the MSSM is the ``Beyond the MSSM'' (BMSSM)
scenario~\cite{Dine:2007xi}.
Here, in addition to the MSSM superpotential a non-renormalizable contribution to the Higgs sector is included
\beq\label{eq:BMSSM1}
W_{\textrm{BMSSM}} = W_{\textrm{MSSM}} +
\frac{\lambda}{M} \, (H_u \, H_d)^2 \; ,
\eeq
as well as a contribution to the soft SUSY-breaking Lagrangian
\beq\label{eq:BMSSM2}
\mathcal{L}_{\textrm{soft}}^{\textrm{BMSSM}} =
\mathcal{L}_{\textrm{soft}}^{\textrm{MSSM}} + \frac{\lambda_s \,
  m_{\textrm{SUSY}}}{M} \, (H_u \, H_d)^2 \; ,
\eeq
where $M$ and $m_\text{SUSY}$ are the energy scale of new physics and the energy scale
of SUSY breaking, respectively.
The new terms in the previous equations account for the leading supersymmetric and $F$-term supersymmetry
breaking corrections to the Higgs sector from a new
threshold at mass scale
$M$\,~\cite{Strumia:1999jm,Casas:2003jx,Brignole:2003cm,PospelovRitz,Dine:2007xi,Antoniadis,Batra:2008rc}\,.
The possible ultraviolet completions can vary significantly, being the simplest the addition of a singlet scalar field,
as in the case of the NMSSM \cite{RevNMSSM}.
It is worth noticing that with the parametrization of Eqs.~\eqref{eq:BMSSM1} and \eqref{eq:BMSSM2},
we can capture the main effects in the effective theory allowing a model-independent description of
a large class of extensions of the MSSM, irrespective of the specific UV completion.
To parametrize the corrections that modify the spectrum and interactions we use the
dimensionless parameters
\beq
\epsilon_{1} \equiv \frac{\lambda \,
  \mu^*}{M} \; , \qquad \epsilon_{2} \equiv \, - \, \frac{\lambda_s \,
  m_{\textrm{SUSY}}}{M} \; .
\eeq
For $M$ of the order of a few TeV, the non-renormalizable corrections of the BMSSM are sizeable and they have been analyzed in the context of electroweak baryogenesis~\cite{Blum:2008ym,Bernal:2009hd,Blum:2010by},
dark matter~\cite{Cheung:2009qk,Bernal:2009hd,Berg:2009mq,Bernal:2009jc,Bae:2010hr,Cassel:2011zd}, Higgs phenomenology~\cite{Carena:2009gx} and collider phenomenology \cite{BMSSMcoll}.
In a nutshell, it has been shown that the consequence of this simple extension of the MSSM including higher
order operators which are suppressed by inverse powers of a scale of
new physics somewhat higher than the electroweak breaking scale,
can  significantly relax the constraints coming from the above mentioned
physical observables. 

In this work, we focus on the implications  of the BMSSM framework when
considering flavour physics observables. We do this in particular for three specific
scenarios:
a generic choice of BMSSM parameters for which EWBG is viable,
and two general frameworks inspired in models where the supersymmetric parameters are correlated
such as the constrained supersymmetric extension of the Standard Model (CMSSM) scenario~\cite{mSUGRA}
and models with non-universal Higgs masses (NUHM)~\cite{Ellis:2002wv}.

In the next section we briefly recall the main features of the model in the Higgs sector.
Section~\ref{sec:flavour} summarizes the main flavour
observables and their current experimental values
that we will use in our analysis.
In Section~\ref{sec:ewb} we briefly present the main effects from
having a viable electroweak baryogenesis scenario
in the BMSSM, which defines the interesting region in
parameter space. Finally in section~\ref{sec:results} we present the main
results of this work which describes the constraints from flavour observables
for the different parametrizations of the BMSSM
that we have considered: the region for which EWBG is viable and two frameworks inspired
by the CMSSM and the NUHM models.

\section{The Higgs boson sector}
\label{sec:spectrum}

\subsection{The spectrum}

We define the scalar Higgs components by
\beq
H_d&=&\begin{pmatrix} H_d^0\\ H_d^- \end{pmatrix}
=\begin{pmatrix}\frac{\phi_1+H_{dr}+iH_{di}}{\sqrt2}\\ H_d^-
\end{pmatrix}\no\\
H_u&=&\begin{pmatrix} H_u^+\\ H_u^0 \end{pmatrix}
=\begin{pmatrix}H_u^+\\ \frac{\phi_2+H_{ur}+iH_{ui}}{\sqrt2}
\end{pmatrix}\;.
\eeq
The  vacuum expectation values  of these Higgs fields are parameterized by
\beq
\langle H_d^0\rangle&=&\phi_1/\sqrt{2}\;,\ \ \ \langle
H_u^0\rangle=\phi_2/\sqrt{2}\;,\\
\tan\beta&=&|\phi_2/\phi_1|\;,\ \ \ v=\sqrt{(\phi_1^2+\phi_2^2)/2}\simeq
174\text{ GeV}\;.\no
\eeq
To leading order, the two charged and four neutral Higgs mass
eigenstates are related to the interaction eigenstates via
\beq\label{eq:Hdef}
\begin{pmatrix} H_d^{*+}\\ H_u^+ \end{pmatrix}&=&
\begin{pmatrix} s_\beta & -c_\beta \\ c_\beta & s_\beta \end{pmatrix}
\begin{pmatrix} H^+ \\ G^+ \end{pmatrix}\;,\no\\
\begin{pmatrix} H_{di}\\ H_{ui} \end{pmatrix}&=&
\begin{pmatrix} s_\beta & -c_\beta \\ c_\beta & s_\beta \end{pmatrix}
\begin{pmatrix} A \\ G^0 \end{pmatrix}\;,\no\\
\begin{pmatrix} H_{dr}\\ H_{ur} \end{pmatrix}&=&
\begin{pmatrix} c_\alpha & -s_\alpha \\ s_\alpha & c_\alpha
\end{pmatrix}
\begin{pmatrix} H \\ h \end{pmatrix}\;,
\eeq
where $c_\beta\equiv\cos\beta$, $s_\beta\equiv\sin\beta$, and
similarly for $\alpha$. Within the MSSM (without the $\epsilon_i$
operators), the angle $\alpha$ is given (at tree level) by 
\beq\label{alphamssm} s_{2\alpha}=-\frac{m_A^2+m_Z^2}{m_H^2-m_h^2}\
s_{2\beta}\;. \eeq 
If the $\epsilon_{1,2}$ couplings are complex, then
the four neutral mass eigenstates are related by a $4\times 4$
transformation matrix to the real and imaginary components of $H_d^0$
and $H_u^0$. In the unitary gauge,
the Goldstone fields $G^\pm$ and $G^0$ are set to zero.

The main effects of the non-renormalizable operators of Eqs.~\eqref{eq:BMSSM1}
and \eqref{eq:BMSSM2} appear on the Higgs masses.
Taking the $Z$ boson mass $m_Z$, the pseudoscalar Higgs boson mass $m_A$ and the ratio of
the two vacuum expectation values $\tan\beta$ as input parameters, we obtain the
leading order corrections in $\epsilon_i$ to the Higgs spectrum:
\beq\label{eq:delmh}
\delta_\epsilon m_h^2&=&2\,v^2\left(\epsilon_{2r}-2\,\epsilon_{1r}\,s_{2\beta}
-\frac{2\,\epsilon_{1r}(m_A^2+m_Z^2)\,s_{2\beta}
+\epsilon_{2r}(m_A^2-m_Z^2)\,c^2_{2\beta}}
{\sqrt{(m_A^2-m_Z^2)^2+4\,m_A^2\,m_Z^2\,s^2_{2\beta}}}\right)\;,
\eeq
\beq
\delta_\epsilon m_H^2&=&2\,v^2\left(\epsilon_{2r}-2\,\epsilon_{1r}\,s_{2\beta}
+\frac{2\,\epsilon_{1r}(m_A^2+m_Z^2)\,s_{2\beta}
+\epsilon_{2r}(m_A^2-m_Z^2)\,c^2_{2\beta}}
{\sqrt{(m_A^2-m_Z^2)^2+4\,m_A^2\,m_Z^2\,s^2_{2\beta}}}\right)\;,
\eeq
\beq
\delta_\epsilon m_{H^\pm}^2&=&2\,v^2\,\epsilon_{2r}\;,
\eeq
where $\epsilon_{1r}$ and $\epsilon_{2r}$ are the real parts of $\epsilon_{1}$ and $\epsilon_{2}$.

The mixing angle $\alpha$ is shifted from its MSSM value:\\
\beq\label{eq:alpha}
s_{2\alpha}&=&\frac{-(m_A^2+m_Z^2)s_{2\beta}+4v^2\epsilon_{1r}}
{(m_H^2-m_h^2)s_{2\beta}}\\
&=&-\frac{(m_A^2+m_Z^2)s_{2\beta}}
{(m_A^4-2m_A^2m_Z^2c_{4\beta}+m_Z^4)^{1/2}}
-4v^2c_{2\beta}^2\frac{2\epsilon_{1r}(m_A^2-m_Z^2)^2
-\epsilon_{2r}s_{2\beta}(m_A^4-m_Z^4)}
{(m_A^4-2m_A^2m_Z^2c_{4\beta}+m_Z^4)^{3/2}}\;.\no
\eeq\\

In the MSSM, the Higgs boson mass can only be larger than the lower bound
given by experimental constraints when large radiative corrections essentially arising from
the stops are invoked: at least one of the stop mass
eigenstates should be rather heavy and/or left-right-stop mixing should be substantial.
On the other hand,  in the BMSSM the additional operators
contribute at tree-level to the lightest Higgs boson mass and diminish
the tensions associated with the stop sector and $\tan\beta$. 
Now, the bound on the Higgs  boson mass allows for stops that are relatively light and unmixed.

\subsection{The vacuum}
 
Previous work has shown that in a generic extension of the MSSM with additional non-renormalizable interactions
two types of vacua can exist, the MSSM-like vacua
and the `supersymmetric electroweak symmetry breaking' (sEWSB) vacua~\cite{Batra:2008rc}. There are significant differences amongst them and therefore the phenomenology both in 
the Higgs sector and other sectors of this supersymmetric model can vary considerably.
Specifically for the present study we perform here we will restrict ourselves to the
MSSM-like vacua and carefully check for the (meta)stability of the vacuum at each
point in parameter space as we scan.
To ensure this the couplings $\epsilon_1$ and $\epsilon_2$ cannot take on arbitrary values. 

The inclusion of the  BMSSM operators may destabilize the scalar potential. If $4|\epsilon_1|>\epsilon_2$, the effective
quartic coupling along one of the D-flat directions is negative, causing a remote vacuum
to form in the presence of which the electroweak  MSSM-like vacuum could become metastable. We make the conservative assumption that when
considering values of $\epsilon_1\gtrsim -0.1$, vacuum stability is ensured provided that the following
condition is fulfilled \cite{Blum:2009na}:
\begin{equation}\label{vacuum}
\frac{m_A^2\,(1+\sin\,2\beta)}{|\mu|^2}\geqslant 2\left(\frac{\tilde\epsilon}{\epsilon_1}\right)^2\left[1+\frac{m_Z^2}{m_A^2}\frac{16\,\tilde\epsilon}{g_Z^2}\left(\frac{1+2\,\sin\,2\beta}{1+\sin\,2\beta}-\frac32\frac{\epsilon_1}{\tilde\epsilon}\right)\right]^{-1}\,,
\end{equation}
where $\tilde\epsilon=\frac14\epsilon_2+\epsilon_1$ and $g_Z^2=g^2+g'^2$ with $g$ and $g'$ the SM gauge couplings.

\begin{figure}[!t]
\begin{center}
\includegraphics[width=8cm]{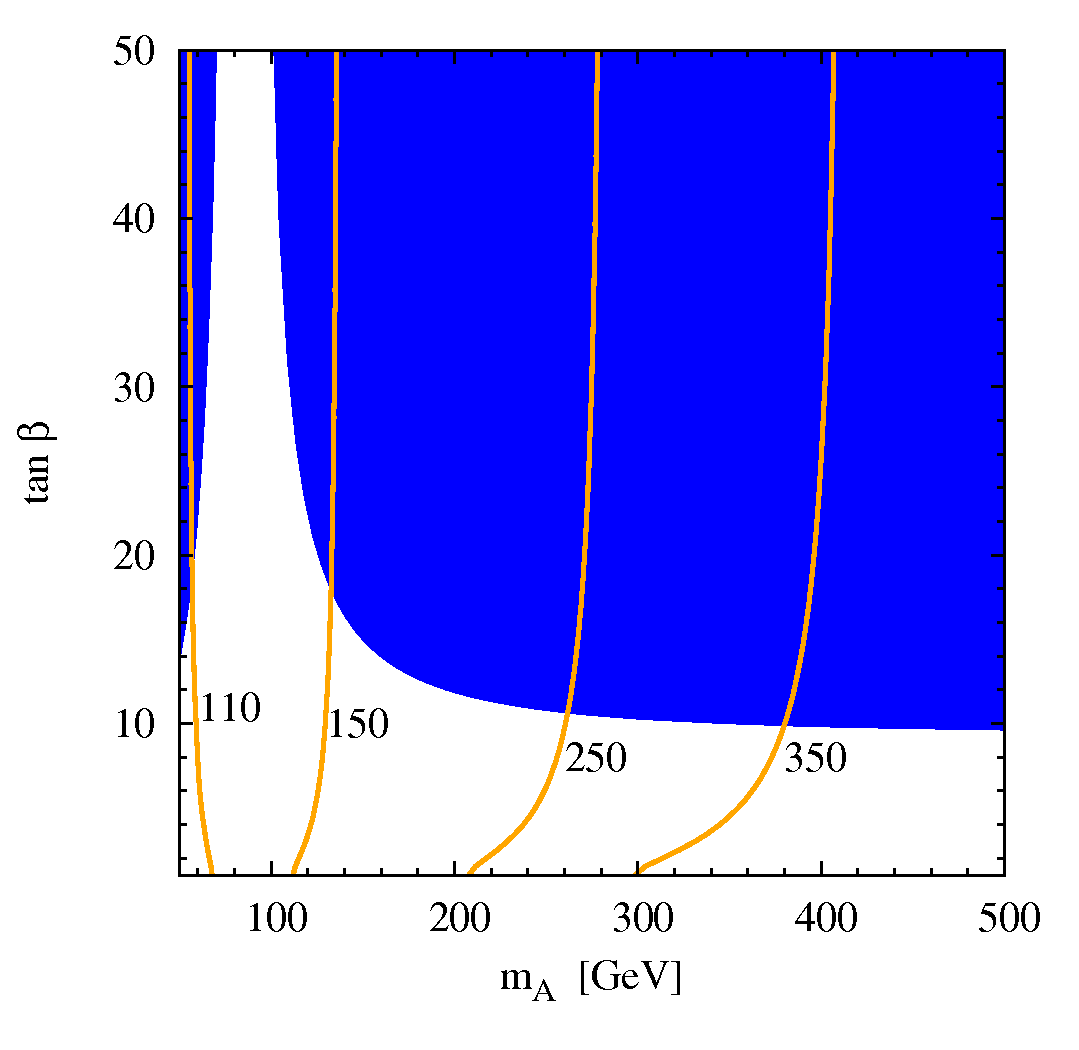}
\end{center}
\vspace{-1cm}
\caption{Regions in the $(m_{A},\tan\beta)$ plane ruled out by the LEP bound on the Higgs mass (blue).
The regions on the right hand side of the orange lines generate an unstable vacuum and are therefore excluded.
These lines correspond to $\mu=110$, $150$, $250$ and $350$ GeV.
$\epsilon_1$ and $\epsilon_2$ are set to -0.1 and 0.05 respectively.\label{tbmA}}
\end{figure}

This will be the condition that we will apply below in the scans.
We illustrate in figure~\ref{tbmA}, the  region in the $\tan\beta$ vs $m_A$,
having fixed the values of $\epsilon_1 =-0.1$ and $\epsilon_2 =0.05$ for which the
MSSM-like vacuum is (meta)stable.
The regions of the parameter space ruled out by the LEP bound on the Higgs mass
are denoted in blue. Both the tree level contribution and the correction coming from
the dimension-5 operators (c.f.~Eq.~\eqref{eq:delmh}) are taken into account.
High values for $\tan\beta$ are usually excluded; only for values around $m_A\sim m_Z$,
higher values for $\tan\beta\gtrsim 10$ are allowed.
The orange lines correspond to contour levels for the vacuum constraint and
for different values of the $\mu$ parameter: $110$, $150$, $250$ and $350$ GeV.
For each value of $\mu$, the regions on the right hand side  of the  contour lines are excluded by the vacuum
stability constraint described in Eq.~\eqref{vacuum}.

\section{Flavour constraints in the MSSM}
\label{sec:flavour}

Flavour physics observables are very sensitive to new physics effects and can play an important role in disentangling different scenarios. In particular, they have been studied extensively and severe constraints have been obtained on the parameters of the MSSM \cite{Carena:2006ai,Heinemeyer:2008fb,Eriksson:2008cx}. Here we compute the most constraining flavour observables with the SuperIso program \cite{Mahmoudi:2007vz,Mahmoudi:2008tp}.

The transition which is most often discussed in this context is the flavour changing neutral current process $b\to s\gamma$ \cite{Ellis:2007ss,Mahmoudi:2007gd,Ellis:2007fu}. Since this transition occurs first at one-loop level in the SM, the new physics contributions can be of comparable magnitude.

The latest combined experimental value for this branching ratio is reported by the Heavy Flavor Averaging Group (HFAG) \cite{Barberio:2008fa}:
\begin{equation}
{\rm BR}(\bar{B} \to X_s \gamma)^{\rm{exp}}=(3.55\pm0.24\pm0.09)\times10^{-4}\,.
\end{equation}

Following \cite{Misiak:2006zs,Misiak:2006ab}, we calculate this branching ratio at the NNLO accuracy. With the most up-to-date parametric inputs as given in \cite{Nakamura:2010zzi} the SM prediction reads\footnote{The slight difference compared to earlier published results is explained by the parametric updates.}:
\begin{equation}
{\rm BR}(\bar{B} \to X_s \gamma)^{\rm{SM}} = (3.06\pm0.22)\times10^{-4} \,.
\end{equation}

The allowed range at 95\% C.L. for this branching ratio, including both the theoretical and experimental uncertainties is \cite{Mahmoudi:2008tp}:
\begin{equation}
2.16\times 10^{-4}\leq {\rm BR}(\bar{B} \to X_s \gamma) \leq 4.93\times 10^{-4}\,.
\end{equation}

In contrast to the $b \to s\gamma$ transitions, the process $B_u\to\tau\nu_\tau$ is sensitive to the charged Higgs boson already at  tree level. Since this decay is helicity suppressed in the SM, whereas there is no such suppression for the charged Higgs boson exchange, these two contributions can be of similar magnitude in the limit of high $\tan\beta$ \cite{Hou:1992sy,Akeroyd:2003zr}. This decay is thus very sensitive to the charged Higgs boson and provides important constraints. 

The current HFAG value for ${\rm BR}(B_u \to \tau\nu_\tau)$ is \cite{Barberio:2008fa} 
\begin{equation}
{\rm BR}(B_u \to \tau\nu_\tau)^{\rm{exp}} = (1.64\pm 0.34)\times 10^{-4}\;.
\end{equation}
The evaluation of BR$(B_u\to\tau\nu_\tau)$ suffers however from the uncertainties in the determination of $|V_{ub}|$. We consider the following ratio to express the new physics contributions:
\begin{equation}
R_{\tau\nu_\tau}=\frac{{\rm BR}(B_u\to\tau\nu_\tau)^{\rm{NP}} }{{\rm BR}(B_u\to\tau\nu_\tau)^{\rm{SM}}} = \left[1-\left(\frac{m_B^2}{M_{H^+}^2}\right)\frac{\tan^2\beta}{1+\epsilon_0\tan\beta}\right]^2 \;.\label{btaunu2}
\end{equation}
In the SM, $R_{\tau\nu_\tau}^{\rm{SM}} = 1$. The experimental result for this ratio is \cite{Barberio:2008fa}:
\begin{equation}
R^{\rm{exp}}_{\tau\nu_\tau}= 1.63\pm 0.54\; ,
\end{equation}
leading to the following allowed interval at 95\% C.L. \cite{Mahmoudi:2008tp}:
\begin{equation}
0.56 < R_{\tau\nu_\tau} < 2.70 \;.
\end{equation}

The semileptonic decays $B \to D\ell\nu$ \cite{Grzadkowski:1991kb,Nierste:2008qe,Kamenik:2008tj} have the advantage of depending on $|V_{cb}|$, which is known to better precision than $|V_{ub}|$. In addition, the $\mathrm{BR}(B \to D \tau \nu_\tau)$ is about $50$ times larger than $\mathrm{BR}(B _u\to \tau \nu_\tau)$ in the SM. Due to the presence of at least two neutrinos in the final state, the experimental determination remains however very complex.
To reduce some of the theoretical uncertainties, we consider the following ratio \cite{Kamenik:2008tj}:
\begin{equation}
\xi_{D\ell\nu} \equiv \frac{{\rm BR}(B \to D \tau \nu_\tau)}{{\rm BR}(B \to D e \nu_e)}\;.
\end{equation}
The SM prediction for this ratio is \cite{Mahmoudi:2008tp}
\begin{equation}
\xi_{D\ell\nu}^{\rm{SM}}=(29\pm 3)\times 10^{-2}\;,
\end{equation}
and the experimental result by the BaBar collaboration reads \cite{Aubert:2007dsa} 
\begin{equation}
\xi_{D\ell\nu}^{\rm{exp}} = (41.6 \pm 11.7 \pm 5.2) \times 10^{-2}\;.
\end{equation}
The 95\% C.L. allowed interval is given by \cite{Mahmoudi:2008tp}
\begin{equation}
0.151 < \xi_{D \ell \nu} < 0.681\;.
\end{equation}
In analogy to the case for $B_u\to \tau\nu_\tau$, charged Higgs bosons would also contribute to the decays  $D_s \to \tau\nu_\tau$ at tree level \cite{Hou:1992sy,Hewett:1995aw,Akeroyd:2003jb,Akeroyd:2007eh,Akeroyd:2009tn}.
The experimental results for this branching ratio is \cite{Barberio:2008fa,Akeroyd:2009tn}:
\begin{equation}
{\rm BR}(D_s \to \tau\nu_\tau)^{\rm{exp}}=(5.38\pm 0.32)\times 10^{-2}\;, 
\end{equation}
while the SM prediction reads:
\begin{equation}
{\rm BR}(D_s \to \tau \nu_\tau)^{\rm{SM}}=(5.10 \pm 0.13) \times 10^{-2}\;, 
\end{equation}%
in which $f_{D_s}= 248 \pm 2.5$ MeV \cite{Davies:2010ip} is used. We consider the following allowed interval at 95\% C.L.:
\begin{equation}
4.7 \times 10^{-2} < {\rm BR}(D_s \to \tau \nu_\tau) < 6.1 \times 10^{-2}\;.
\end{equation}
The last leptonic decay that we consider is the decay $K \to \mu \nu_\mu$, and in particular we consider the ratio \cite{Antonelli:2008jg}
\begin{equation}
R_{\ell 23}=\left|1-\frac{m^2_{K^+}}{M^2_{H^+}}\left(1 - \frac{m_d}{m_s}\right)\frac{\tan^2\beta}{1+\epsilon_0\tan\beta}\right|\;.\label{kmunu2}
\end{equation}
The SM prediction for this ratio is $R_{\ell 23}^{\rm{SM}}=1$ and the experimental measurement gives:
\begin{equation}
R^\mathrm{exp}_{\ell 23}=1.004 \pm 0.007\;,
\end{equation}
where $f_K/f_\pi = 1.189 \pm 0.007$ is used \cite{Follana:2007uv}, and the allowed interval at 95\% C.L. read:
\begin{equation}
0.990 < R_{\ell 23} < 1.018\;.
\end{equation}
In the following sections, we study the regions excluded in the BMSSM parameter space by the flavour observables.

\section{EWBG in the BMSSM Framework}
\label{sec:ewb}

Electroweak baryogenesis (EWBG) is an attractive mechanism for
generating the baryon asymmetry of the Universe (BAU), in part due to its
testability.  However, for this mechanism to be successful in the MSSM the
parameter space is reduced to a finely tuned region.

In the stop sector  of the MSSM a strong
first-order phase transition requires at least one light stop (which
must be mostly right-handed ($\widetilde t_R$), to avoid large contributions to the
$\rho$ parameter and due to null searches for a light
sbottom~\cite{Drees:1990dx,Nakamura:2010zzi}).  At the same time, the
large radiative corrections needed to increase the Higgs boson mass above
the LEP bound $m_h>114$ GeV~\cite{Nakamura:2010zzi} require that at
least one stop ($\widetilde t_L$) is very heavy \cite{Quiros:2000wk}.
The electroweak phase transition was studied in an effective
theory with a large stop hierarchy, concluding that successful EWBG is possible only for
$m_{\tilde t_R}<125$~GeV and $m_{\tilde t_L}>6.5$~TeV~\cite{Carena:2008vj}, making the scenario finely tuned.

More recently, it was shown \cite{Blum:2008ym,Bernal:2009hd} that in the BMSSM the left-handed stop can also be relatively light, providing additional
bosonic degrees of freedom that strengthen the first-order phase
transition. The $\rho$ parameter and direct searches for a light sbottom are
now the main constraint.

On the other hand, large values of  the pseudo-scalar Higgs boson,
 $m_A$, are preferred (i)~to make the electroweak phase transition~\cite{brignole,Espinosa:1996qw,Carena:1996wj,all,Carena:2008vj} more
strongly first-order, and (ii) to evade constraints from $b \to s
\gamma$~\cite{Cirigliano:2009yd}.  However, the production of
left-handed charge during EWBG is enhanced when $m_A$ is light.

In the BMSSM, the pseudoscalar Higgs boson can be significantly lighter as the phase transition
is strengthened by having lighter stops.

The value of $\tan\beta$ is limited  from a compromise in
giving a large enough value of the Higgs mass versus a strong enough
phase transition, and in contrast from the constraints from $b \rightarrow s
\gamma$ for small values of $m_A$.

As far as the EWBG in the BMSSM,  given that the Higgs boson mass is made large enough
from the additional corrections from the non-renormalizable operators, $\tan\beta$ can take on smaller values. Let us recall that large values for $\tan\beta$ decrease the contribution to the Higgs boson mass
coming from the dimension 5 operators, and in general values of $\tan\beta\gtrsim 10$ tend to be unfavourable.
However, for $m_A\sim m_Z$ the uplift of the Higgs boson mass is maximal and therefore $\tan\beta$
could take much higher values.

\subsubsection*{CP violation for BAU production}
In the MSSM, the CP-violating phases that drive EWBG arise in the
gaugino/higgsino sector and at the same time contribute to electric dipole moments (EDMs). 
One-loop contributions can be sufficiently suppressed by making the
first two squark and slepton generations heavy,  however, there exist two-loop
contributions that cannot be suppressed without spoiling EWBG
 and give a minimum value of the EDM.  The main conclusion is that EWBG
 with universal gaugino phases is nearly ruled out. With
improvements by a factor $3\!-\!4$ in the upper bounds on the EDMs of
the electron or the neutron, MSSM baryogenesis will be possible only
in the so-called ``bino-driven'' scenario, where the CP-violating
phase associated with the U(1)$_Y$ gaugino is tuned to be much larger
than that of the SU(2)$_L$ gaugino~\cite{Li:2008ez,Cirigliano:2009yd}.

In contrast in the BMSSM, new phases arise that can produce the baryon asymmetry of the Universe \cite{Blum:2010by} and relax the constraints on MSSM parameters. In fact, the possibility of spontaneous baryogenesis determined from the phase in the Higgs sector must be taken into account. Furthermore, the top and stop CP-violating sources are now unsuppressed in comparison to the MSSM case.
In fact the MSSM phases can be zero and still EWBG is viable, however, if the experimental sensitivity of electric dipole moment experiments is increased by one order of magnitude in the BMSSM scenario a EDM signal should be detected.
 
The imaginary parts of $\epsilon_1$ and $\epsilon_2$ are constrained from EDMs~\cite{Blum:2010by}.
However, we will for simplicity take them to be zero in this work, otherwise the
Higgs scalar sector is significantly affected and a more complicated scenario arises.

\section{Results}
\label{sec:results}

For the numerical evaluation we use SuperIso v2.8 \cite{Mahmoudi:2007vz,Mahmoudi:2008tp}, and the spectra of SUSY particles are calculated using a modified version of SuSpect \cite{Djouadi:2002ze,Bernal:2009hd} incorporating the leading order corrections in $\epsilon_i$ to the Higgs masses and to the mixing angle $\alpha$. The top quark pole mass is set to $m_t = 173.3$ GeV \cite{:1900yx}.

For each observable we determine the regions excluded in the BMSSM parameter space. We perform the scans in the following way:
we first impose the vacuum stability constraint, the constraint on the value of the lightest Higgs boson mass, the constraint from fulfilling the $b\rightarrow s\gamma$ branching ratio, followed by the rest of the flavour constraints.

To understand correctly the plots below, it is important to keep in mind that we are performing a multiple-parameter scan, as such
when projected onto a two-dimensional plane all other parameters can take on many different values.

\subsection{Flavour constraints on EWBG}

We present our first results in the case in which we fix the parameters in the BMSSM such that  the electroweak baryogenesis
scenario is feasible. As mentioned above we do not consider in this work the effect of complex values for $\epsilon_1$ 
and $\epsilon_2$.
Based on the discussion presented in previous sections we fix the parameters of the effective theory as follows:
we take the gaugino masses to be
$M_1 =100$ GeV, $M_2 =200$ GeV and $M_3= 1000$ GeV.
The right handed stop $m_{\tilde{t}_{R}} = 160$ GeV,
for the left-handed  supersymmetry breaking stop mass we scan in the range  $150 < m_{Q_{3}} < 300$ GeV.
The Higgsino mass parameter $100< \mu < 400$ GeV.
We assume positive values for the $\mu$ parameter since this makes it easier to explain $(g-2)_\mu$ data
and to satisfy $b\to s\gamma$ constraints.
The trilinear scalar couplings $0<A_t = A_c = A_u < 200$ GeV.
We fix also all other sleptons and squarks soft masses $m_Q=m_L =1000$ GeV.

Recall that the flavour observables that we focus on are essentially:
$b\rightarrow s \gamma$,
$B\rightarrow \tau \nu$,
$R_{l23}$,
$B\rightarrow D\ell \nu$ and
$D_s \rightarrow \tau \nu$.

\begin{figure}[!ht]
\begin{center}
\includegraphics[width=8.4cm]{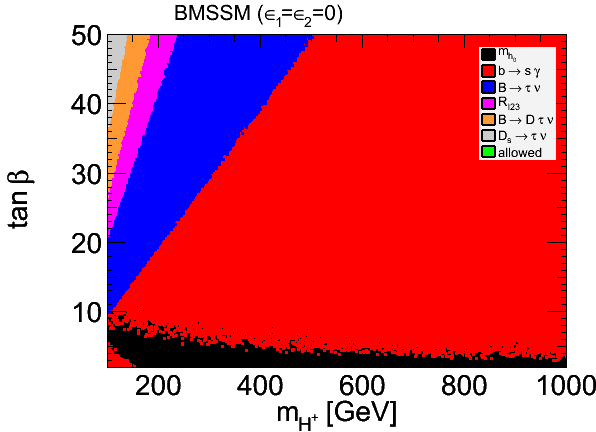}\\
\includegraphics[width=8.4cm]{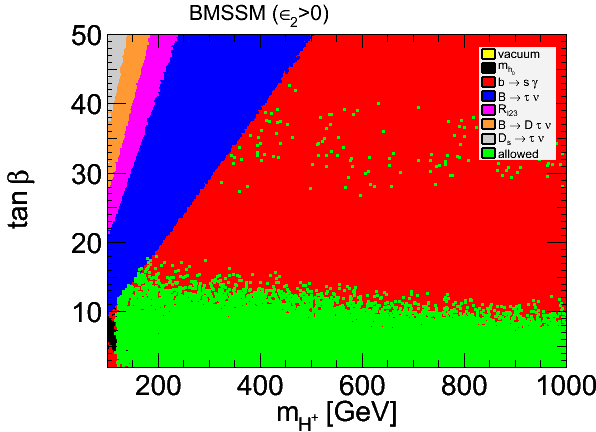}
\includegraphics[width=8.4cm]{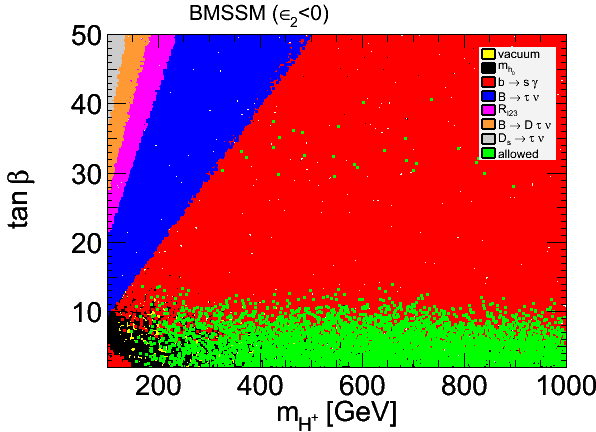}
\end{center}
\caption{Flavour constraints on the $(m_{H^+},\tan\beta)$ plane in the MSSM (upper) and the BMSSM (lower) for the scenario where EWBG is feasible, described in text. The lower left panel corresponds to $\epsilon_2>0$ and the lower right panel to $\epsilon_2<0$.
\label{tbmH}}
\end{figure}

In figure \ref{tbmH} we compare the constraints in the $\tan\beta$ vs $m_{H^{+}}$
plane with (lower panels) and without (upper panel) the terms arising from
the higher dimensional operators.
In the plots, in addition to scanning over the SUSY parameters as described above,
we scan over values of $-0.1 <\epsilon_1< 0$ and $0<\epsilon_2< 0.1$ (lower left panel)
or $-0.1<\epsilon_2<0$ (lower right panel). It is clear that
different flavour observables rule out some of the regions and 
we observe that a wide band, in the range of $\tan\beta \lesssim 15$ for $\epsilon_2>0$ and
$\tan\beta \lesssim 10$ for $\epsilon_2<0$, for $m_{H^+}\gsim 150-200$ GeV,
 that was excluded in the MSSM, can now satisfy all constraints. This occurs given our choices for the values of the stop masses, which in the MSSM need to be very large for the Higgs boson to avoid the experimental limit. In the BMSSM this can be avoided as mentioned above and thus the region would be allowed from flavour physics constraints. There are also some scattered allowed points for  $30<\tan\beta<40$ when
 $m_{H^+}>300$~GeV.
It is important to note that the lowest excluded values of $\tan\beta$, as a function of the charged Higgs boson mass, from the $b\rightarrow s\gamma$ constraint are increased in the BMSSM case as compared to the MSSM, with the constraint being stronger in the case of $\epsilon_2<0$.

\newpage
The vacuum stability and the lightest Higgs boson mass constraint are not visible since they are displayed in the background. 
However, as expected from equation~\eqref{vacuum}, the vacuum constraint tends to be stronger for low values of the charged Higgs mass (implying low $m_A$ values) and for negative values for $\epsilon_2$, and the region with $m_{H^{+}} \lsim 200$~GeV becomes excluded.

We conclude that for the EWBG scenario  only  the regions of low $\tan\beta$  and  $m_{H^+}\gsim 150-200$~GeV  and a few points  above  $\tan\beta=25$ and $m_{H^{+}} \lsim 300$~GeV will remain  viable.

\begin{figure}[!t]
\begin{center}
\includegraphics[width=8.4cm]{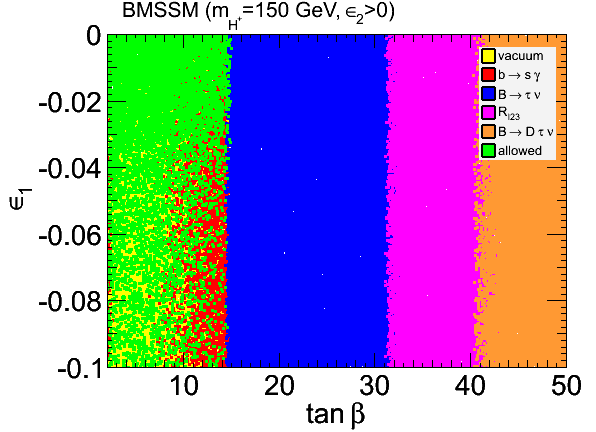}
\includegraphics[width=8.4cm]{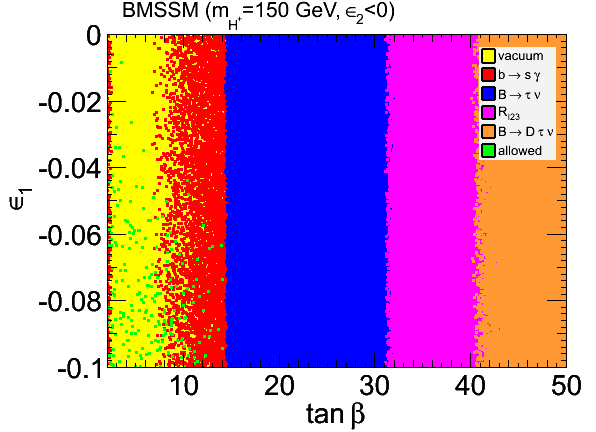}
\end{center}
\caption{Flavour constraints on the $(\tan\beta,\,\epsilon_1)$ plane in the BMSSM for the scenario where EWBG is feasible. We used $m_{H^\pm}=150$ GeV and $\epsilon_2$ positive (left panel) or $\epsilon_2$ negative (right panel).\label{eps1tb}}
\end{figure}

In figure \ref{eps1tb} we show the regions of parameter space that are allowed after applying all constraints;
in this case we are setting the charged Higgs boson mass $m_{H^+}=150$~GeV. 
We do this separately for both $\epsilon_2>0$ (left panel) and $\epsilon_2<0$ (right panel).

In this figure, the flavour observables (and in particular $b\to s\gamma$ and $B\to\tau\nu_\tau$) are quite restrictive leaving the allowed region limited to values of $\tan\beta \lesssim 15$.
Moreover, let us note that the vacuum stability constraint cuts out regions of parameter space that otherwise would be allowed purely from flavour constraints.
In the case of $\epsilon_2<0$, the vacuum stability condition is especially constraining,
limiting the allowed parameter space to values of $\epsilon_1\lesssim-0.05$.
On the other hand, for $\epsilon_2>0$ the vacuum condition tends to be alleviated, and consequently the allowed region satisfying all constraints is defined over the full range of $\epsilon_1$, with a mild dependence on $\tan\beta$.
The different flavour constraints have essentially the same effect for both cases with $\epsilon_2>0$
and $\epsilon_2<0$. 

\begin{figure}[!ht]
\begin{center}
 \hspace*{-1cm}\includegraphics[width=8.4cm]{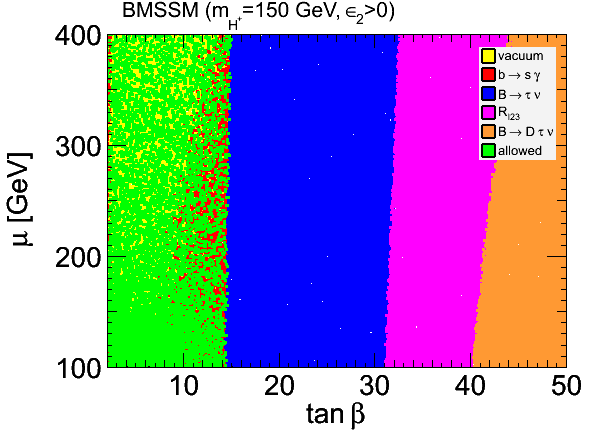}\includegraphics[width=8.4cm]{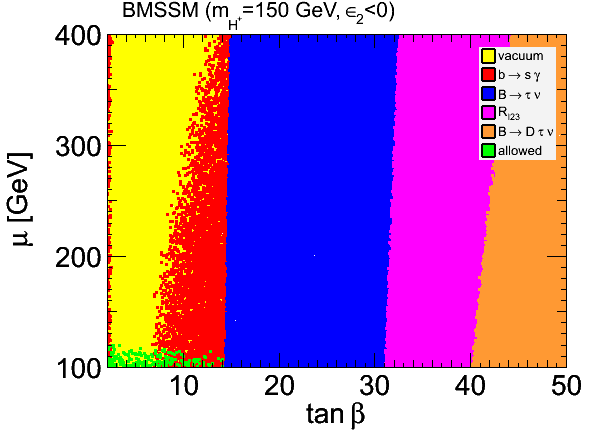}\\
 \hspace*{-1cm}\includegraphics[width=8.4cm]{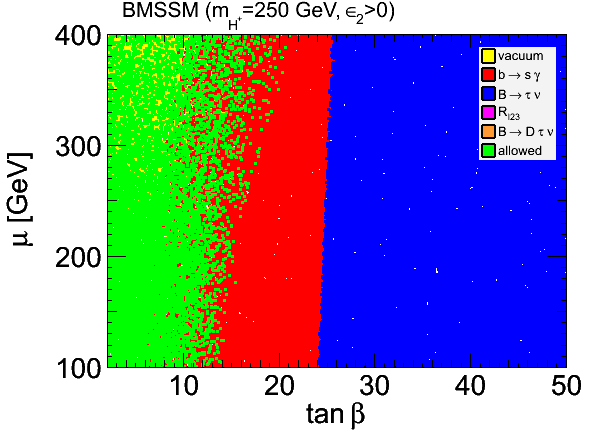}\includegraphics[width=8.4cm]{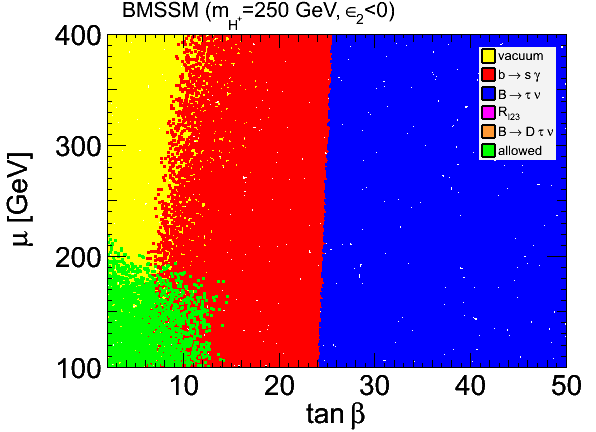}\\
 \hspace*{-1cm}\includegraphics[width=8.4cm]{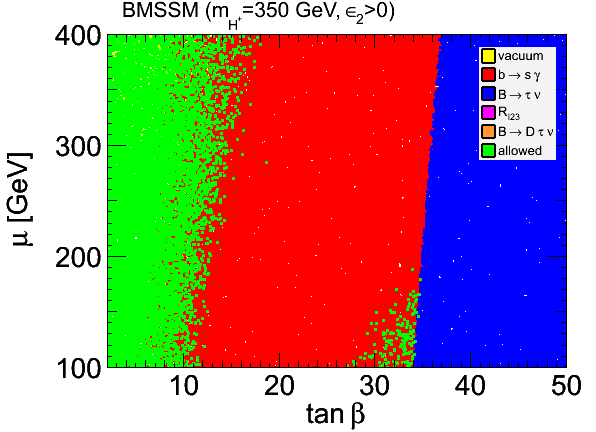}\includegraphics[width=8.4cm]{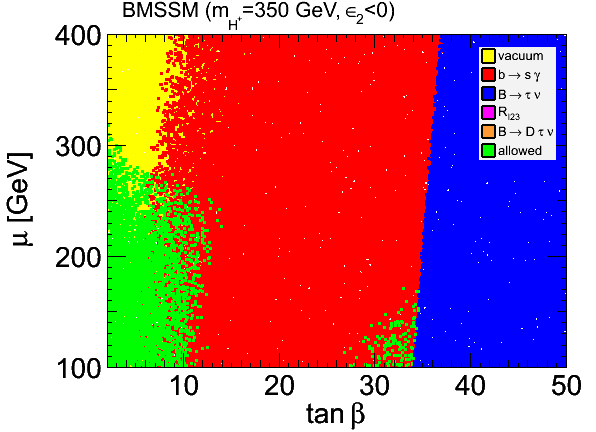}
\end{center}
\caption{Flavour constraints on the $(\tan\beta,\,\mu)$ plane in the BMSSM
for the scenario where EWBG is feasible.
We used $\epsilon_2>0$ for the left panels and $\epsilon_2<0$ for the right panels.
We also set $m_{H^\pm}=150$ GeV (upper panels),
$m_{H^\pm}=250$ GeV (central panels) and $m_{H^\pm}=350$ GeV (lower panels).\label{mutbmH}}
\end{figure}

We next study in detail in figure \ref{mutbmH} the constraining effect from the flavour observables and the vacuum stability constraint in the plane $\mu$ vs
$\tan\beta$ for three different values of the charged Higgs boson mass:
$m_{H^{+}}=150$ (upper panels), $250$ (central panels) and $350$ GeV (lower panels);
and for $\epsilon_2$ positive (left panels) and negative (right panels).

\newpage
In the $\epsilon_2>0$ case with the increase of $m_{H^\pm}$ the allowed parameter space tends to shrink mainly
due to the strengthening of the $b\to s\gamma$ constraint for low values of the $\mu$ parameter.
On the contrary, the bound given by all the other flavour observables and in particular by $B\to\tau\nu_\tau$ weaken.
The vacuum condition is a very weak constraint for positive values of $\epsilon_2$. 
However, for negative values this condition is much stronger.
In this case there is a significant effect in pushing the upper bound on $\mu$ to larger values,
up to $\mu\sim 300$ GeV for $m_{H^\pm}=350$ GeV.
For increasing $m_{H^+}$ values, the constraint from $B\rightarrow \tau\nu_\tau$ also
weakens but still eliminates regions in parameter space with larger values of $\tan\beta$.
This is consistent with the fact mentioned above of the dependence at tree level of
the $B\rightarrow \tau\nu_\tau$ with the charged Higgs boson.
The constraint from $b\rightarrow s\gamma$ still limits the allowed range to $\tan\beta\lesssim 12$. 
Note also that some allowed points appear for $30<\tan\beta<35$, with $\mu\lesssim 200$ GeV when $m_{H^+}=350$ GeV.
Similar effects are also seen in the $\epsilon_2>0$ case. That is the role played by the
constraint from $b\rightarrow s\gamma$ which is highlighted for intermediate values of $\tan\beta$. Note that $D_s \to \tau \nu_\tau$ does not provide any constraint in these examples.

\subsection{Flavour constraints for a CMSSM-like model}

We now consider the impact of the flavour constraints in the case of CMSSM with the addition of the
non-renormalizable operators.
In order to understand these implications of the BMSSM framework and, in particular,
in order to allow for a simple comparison with CMSSM-like models, we investigate
the following framework. The MSSM parameters that we use are those that would have
corresponded to a CMSSM model specified by the five parameters:
$m_0$, $m_{1/2}$, $A_0$, $\tan\beta$ and sign($\mu$).
Thus, the correlations between the low energy MSSM parameters are the same as those
that would hold in a CMSSM framework.
Let us emphasize again that one should {\it not} think about
this set of parameters as coming from an extended CMSSM model, since the effects of
the BMSSM physics at the few TeV scale on the running are not (and cannot be) taken
into account.

\begin{figure}[!t]
\begin{center}
\includegraphics[width=8.4cm]{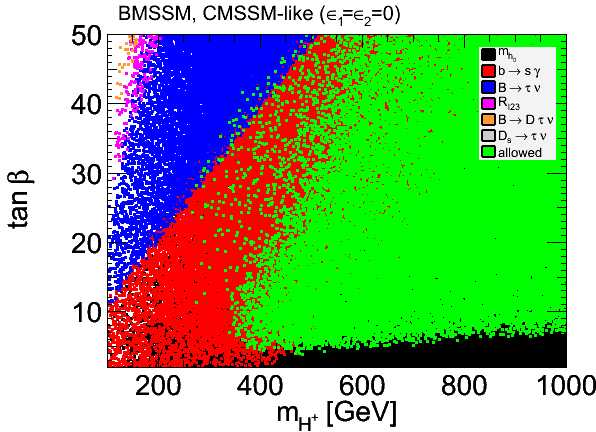}\\
\includegraphics[width=8.4cm]{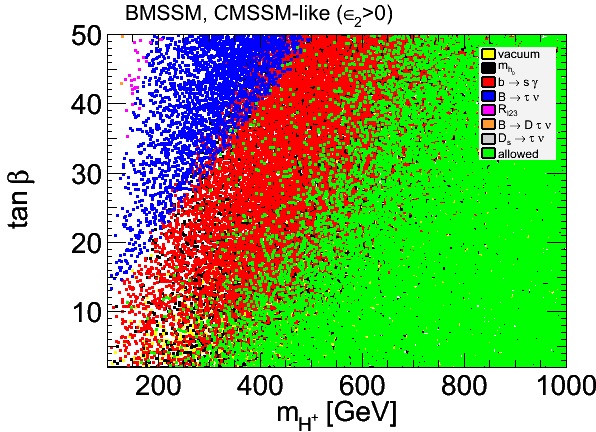}
\includegraphics[width=8.4cm]{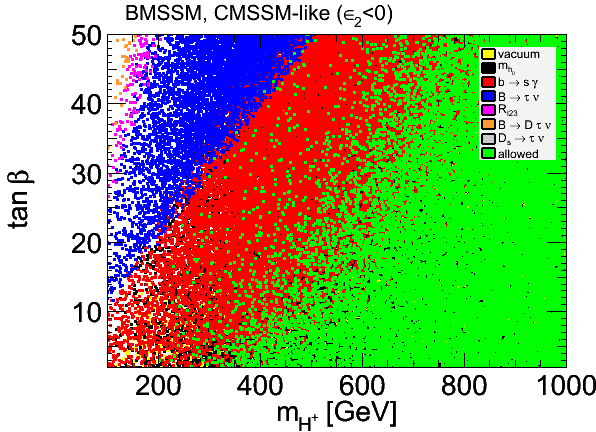}
\end{center}
\vspace{-0.5cm}
\caption{Flavour constraints on the $(m_{H^+},\tan\beta)$ plane for the CMSSM (upper panel)
and the CMSSM-like model with non-renormalizable operators (lower panels).
The lower left panel corresponds to $\epsilon_2>0$ and the lower right panel to $\epsilon_2<0$.\label{CMSSM}}
\end{figure}

In the scans, we took  $-0.1 < \epsilon_1 < 0$ and $|\epsilon_2| <0.1$, and varied the CMSSM parameters in the ranges $50 < m_0 < 1200$ GeV, $100 < m_{1/2} < 1000$ GeV, $-1000 < A_0 < 1000$ GeV, $ 2 < \tan\beta < 50$ and imposed $\mu >0$ for $(g -2)_\mu$ compatibility. 

Figure \ref{CMSSM} shows in the $\tan\beta$ vs $m_{H^+}$ plane a comparison of the allowed regions, and the effect of the different regions that are ruled out by the vacuum, $m_h$, and the flavour physics restrictions.

While in the MSSM case very low values for $\tan\beta$ are ruled out by the constraint on the mass of the lightest Higgs boson,
in the BMSSM model for values of $m_{H^+} \gsim 300$ GeV, this region is once again viable.
Another important difference, is the effect of the $m_h$ constraint for the MSSM in region in which $2 < \tan\beta < 50$, and $150 < m_{H^+} < 500$ GeV which essentially rules out a large portion except for large values of $\tan\beta$,  while in the BMSSM, it is mostly the vacuum constraint and $b\rightarrow s\gamma$ which exclude this area, in particular for the case with $\epsilon_2<0$. 

The  comparative effect  for both models from the other flavour observables: $B\rightarrow \tau \nu$,  $R_{l23}$,
$B\rightarrow D\ell \nu$, $D_s \rightarrow \tau \nu$ is rather mild.

\subsection{Flavour constraints for a NUHM-like model}

We explore in this section the parameter space of the non-universal Higgs mass (NUHM)
framework~\cite{Ellis:2002wv},
in which the universality assumptions of the soft SUSY breaking contributions to the
Higgs masses are relaxed as compared to the CMSSM scenario. Within this framework, two
additional free parameters, $m_A$ and $\mu$, add to the five universal parameters of the CMSSM
scenario. Usually one trades $m_A$ for $m_{H^+}$ through the mass relation, and therefore the charged Higgs boson mass can be treated essentially as a free parameter. This makes this model more attractive for the study of the Higgs sector. We also take into account the two extra parameters $\epsilon_1$ and $\epsilon_2$, in order to parametrize the effects of the non-renormalizable operators.

To study the effect of the different constraints on a NUHM-like scenario, we perform scans over $100 < m_0 < 500$ GeV, $100 < m_{1/2} < 1000$ GeV, $-1000 < A_0 < 1000$ GeV, $100 <~m_A <~1000$ GeV, $100 < \mu <1000$ GeV and $ 2 < \tan\beta < 50$. In addition we scan over $-0.1 < \epsilon_1 < 0$ and $|\epsilon_2| <0.1$.

\begin{figure}[!t]
\begin{center}
\includegraphics[width=8.4cm]{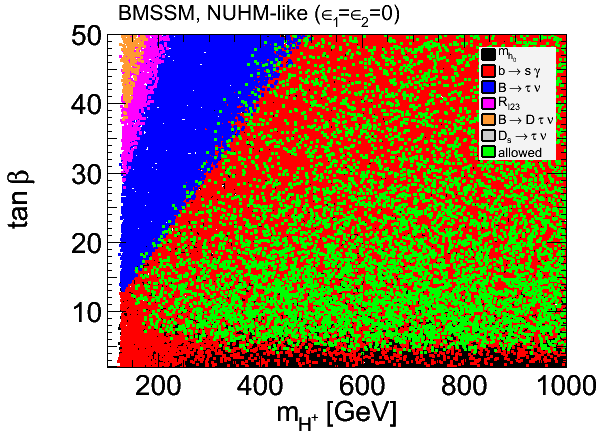}\\
\includegraphics[width=8.4cm]{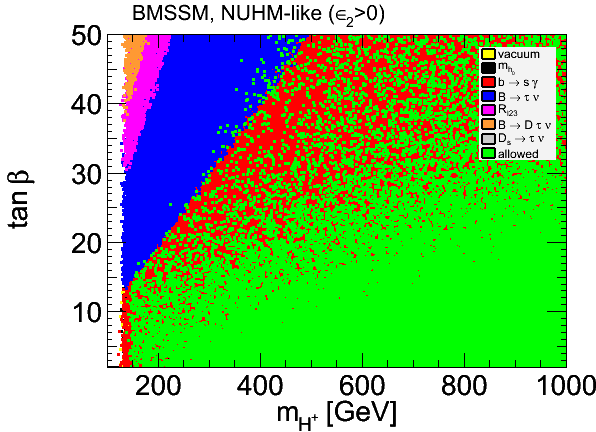}
\includegraphics[width=8.4cm]{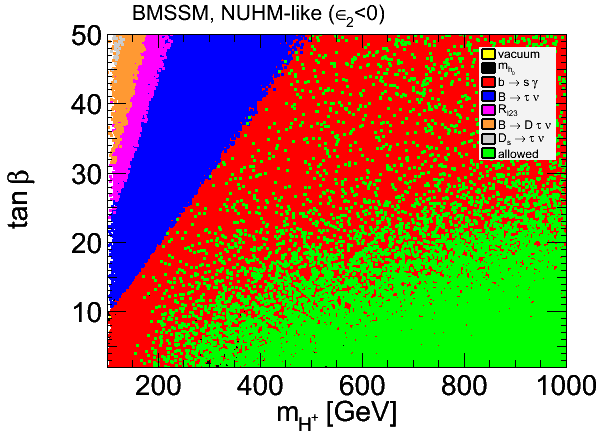}
\end{center}
\vspace{-0.5cm}
\caption{Flavour constraints on the $(m_{H^+},\tan\beta)$ plane for the plain NUHM (upper panel)
and the NUHM-like model with non-renormalizable operators (lower panels).
The lower left panel corresponds to $\epsilon_2>0$ and the lower right panel to $\epsilon_2<0$.\label{tbmHNUHM}}
\end{figure}

We present in figure \ref{tbmHNUHM} the allowed regions in the $\tan\beta$ vs $m_{H^{+}}$ plane which satisfy the constraints from flavour observables and  the effect of including the vacuum stability constraint for both $\epsilon_2>0$ (lower left panel) and $\epsilon_2<0$ (lower right panel).
For completeness we are also showing the plain NUHM case with $\epsilon_1=\epsilon_2=0$ in the upper panel.
In the MSSM limit, very low values of $\tan\beta$ are excluded all across the range of values of $m_{H^+}$
because of the Higgs mass; this region is allowed in the BMSSM for both positive and negative values
of $\epsilon_2$.
While the regions excluded from the constraint arising from $B\rightarrow\tau\nu_\tau$ are roughly similar
for the MSSM and the BMSSM case with $\epsilon_2>0$, in the case of the BMSSM with $\epsilon_2<0$,
the excluded region is somewhat enlarged reaching smaller values of $\tan\beta=10$ for small values of $m_{H^+}$.
In this latter case a more preponderant role is played in excluding regions of parameter space
by the constraints from $b\to s\gamma$ 
and the vacuum stability that rule out almost all the parameter space for $m_{H^\pm}\lesssim 200$~GeV.
Regions with low values of $\tan\beta$ and low $m_{H^{+}}$ are excluded. 

\begin{figure}[!ht]
\begin{center}
\includegraphics[width=8.4cm]{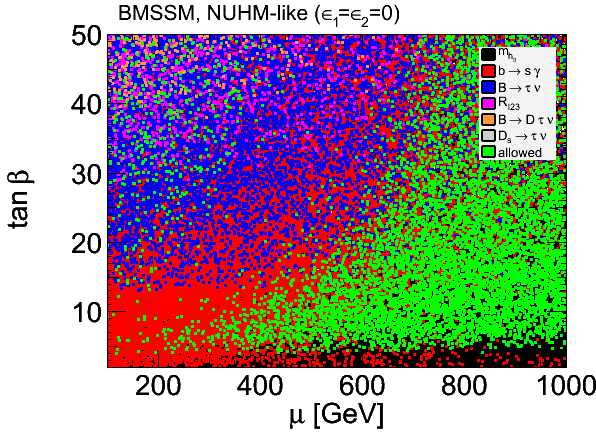}\\
\includegraphics[width=8.4cm]{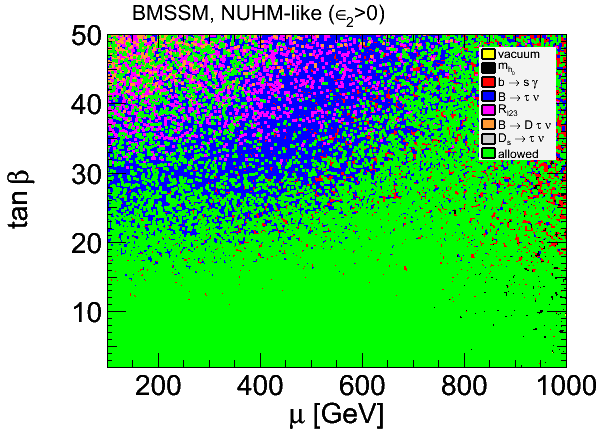}
\includegraphics[width=8.4cm]{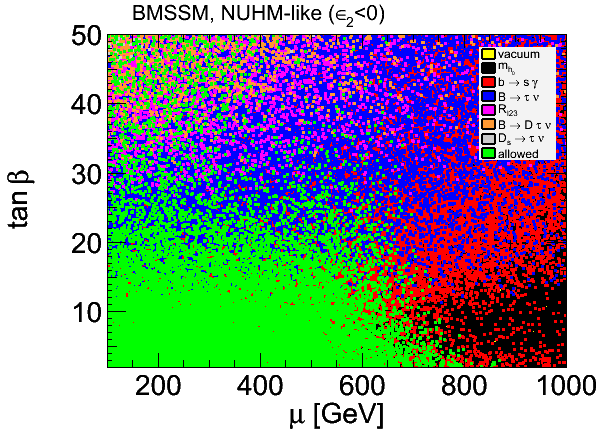}
\end{center}
\vspace{-0.5cm}
\caption{Flavour constraints on the $(\mu,\tan\beta)$ plane for the plain mSUGRA (upper panel)
and the mSUGRA-like model with non-renormalizable operators (lower panels).
The lower left panel corresponds to $\epsilon_2>0$ and the lower right panel to $\epsilon_2<0$.\label{tbmuNUHM}}
\end{figure}

\newpage
Figure \ref{tbmuNUHM} shows the impact of different flavour constraints in the $(\mu,\tan\beta)$
parameter space for $\epsilon_2>0$ (lower left panel) and $\epsilon_2<0$ (lower right panel),
and the standard NUHM case (upper panel).
In this case, the bound on the Higgs boson mass excludes low values of $\tan\beta\lesssim 5$.
For both BMSSM cases, the impact is lifted compared to the NUHM case.
For negative $\epsilon_2$ values, the vacuum stability condition, $b\to s\gamma$ and also $B\rightarrow\tau\nu_\tau$ rule out a large portion
of the parameter space corresponding to $\mu\gtrsim 700$~GeV, compared to both the standard NUHM and the BMSSM with $\epsilon_2>0$.

\section{Conclusions}

We have presented a detailed analysis in the context of the BMSSM model of the effect of constraining the model from the perspective of flavour physics. In our study we have focused on the regions of parameter space in which foremost the vacuum is MSSM-like and sufficiently long-lived. We have then analyzed and compared the possible restrictions of the parameters in the model
from flavour physics observables. We find that there can be significant differences compared to the MSSM case when we apply the flavour constraints.

In summary we find:

\begin{itemize}
\item The flavour constraints on EWBG, especially from $b \rightarrow s \gamma$ and $ B \rightarrow \tau\nu_\tau$ limit the
value of $\tan\beta \lsim 15$ and  furthermore for $\epsilon_2 < 0$, the vacuum stability constraint requires $\epsilon_1 \lsim -0.05$ and strongly restricts the allowed values of $\mu$ depending on the charged Higgs boson mass. For large enough values of  the charged Higgs boson mass the constraint arising from $b \rightarrow s \gamma$ is determinant for intermediate values of $\tan\beta$.
\item The flavour constraints in the model with a CMSSM parametrization show that very low values of $\tan\beta$ are allowed for large enough values of the charged Higgs boson mass. Here the exclusion of large regions in parameter space from the $b\rightarrow s \gamma$ constraint is further strengthened in particular when $\epsilon_2 <0$.
\item Finally in the model we studied with a NUHM-like parametrization, the  $b\rightarrow s \gamma$ constraint has a stronger impact in excluding regions of parameter space especially for small values of the charged Higgs boson mass. In addition for $\epsilon_2 < 0$ values of $\mu \gsim 700$ GeV are ruled out by flavour constraints.
\end{itemize}

\section*{Acknowledgements}
NB is supported by the DFG TRR33 `The Dark Universe'.

\end{document}